\documentclass[conference]{IEEEtran}
\renewcommand\IEEEkeywordsname{Keywords}
\usepackage[T1]{fontenc}
\usepackage{textcomp}% for straight quote in lstlisting
\ifCLASSOPTIONcompsoc
  \usepackage[nocompress]{cite}
\else
  \usepackage{cite}
\fi
\ifCLASSINFOpdf
\else
\fi
\usepackage{amsmath}
\usepackage{amssymb}
\usepackage{array}
\ifCLASSOPTIONcompsoc
  \usepackage[caption=false,font=footnotesize,labelfont=sf,textfont=sf]{subfig}
\else
  \usepackage[caption=false,font=footnotesize]{subfig}
\fi
 \usepackage{dblfloatfix}
\usepackage{url}
\usepackage{listings}
\usepackage[utf8]{inputenc}
\usepackage{enumitem}
\usepackage{todonotes}

\usepackage{csquotes}

\usepackage{balance}

\usepackage{hyperref}
\hypersetup{hidelinks=true}

\newcommand{\Sc}{smart contract}
\newcommand{\SC}{Smart contract}

\newcommand{\tz}{tez}
\lstdefinelanguage{michelson}{%
   columns=fullflexible,%
   basicstyle=\tt ,
   commentstyle=\slshape,%
   keywords={%
     \{,\},
     DROP, DUP, SWAP, PUSH, SOME, NONE, UNIT, IF_NONE,%
   PAIR, CAR, CDR, LEFT, RIGHT, IF_LEFT, IF_RIGHT, NIL,%
   CONS, IF_CONS, SIZE, EMPTY_SET, EMPTY_MAP, MAP, ITER,%
   MEM, GET, UPDATE, IF, LOOP, LOOP_LEFT, LAMBDA, EXEC,%
   DIP, FAILWITH, CAST, RENAME, CONCAT, SLICE, PACK,%
   UNPACK, ADD, SUB, MUL, EDIV, ABS, NEG, LSL, LSR,%
   OR, AND, XOR, NOT, COMPARE, EQ, NEQ, LT, GT, LE,%
   GE, SELF, CONTRACT, TRANSFER_TOKENS, SET_DELEGATE,%
   CREATE_ACCOUNT, CREATE_CONTRACT, CREATE_CONTRACT,%
   IMPLICIT_ACCOUNT, NOW, AMOUNT, BALANCE, CHECK_SIGNATURE,%
   BLAKE, SHA, SHA, HASH_KEY, STEPS_TO_QUOTA, SOURCE,%
   SENDER, ADDRESS,%
   CMPEQ,CMPNEQ,CMPLT,CMPGT,CMPLE,CMPGE,%
   IFEQ,IFNEQ,IFLT,IFGT,IFLE,IFGE,%
   IFCMPEQ,IFCMPNEQ,IFCMPLT,IFCMPGT,IFCMPLE,IFCMPGE,%
   ASSERT,%
   ASSERT_EQ,ASSERT_NEQ,ASSERT_LT,ASSERT_LE,ASSERT_GT,ASSERT_GE,%
   ASSERT_CMPEQ,ASSERT_CMPNEQ,ASSERT_CMPLT,ASSERT_CMPLE,ASSERT_CMPGT,ASSERT_CMPGE,%
   ASSERT_NONE,ASSERT_SOME,%
   ASSERT_LEFT,ASSERT_RIGHT,%
   UNPAIR,%
   },%
   alsoletter={'},
   upquote=true,
   keywordstyle={\color{purple}\sffamily},%
   morekeywords=[2]{%
     key, unit, signature, option, list, set, operation, address,%
     contract, pair, or, lambda, big_map, map,%
     int, nat, string, bytes, mutez, bool, key_hash, %
     timestamp, 'a, 'b, 'S, 'p%
   },%
   keywordstyle=[2]{\color{purple}\ttfamily},%
   classoffset=2,%
   morekeywords=[3]{%
     storage, parameter, code %
   },%
   keywordstyle=[3]{\bfseries},%
   sensitive,%
   comment=[l]\#,%
   morestring=[d]",%"
   literate={->}{{$\rightarrow$}}1%
}[keywords,comments,strings]%

\lstMakeShortInline[language=michelson]!

\lstdefinelanguage{rpcs}{%
   columns=fullflexible,%
   breaklines=true,
   postbreak={\raisebox{0ex}[0ex][0ex]{\hspace{-2.5em}\ensuremath{\hookrightarrow}\qquad}},
   upquote=true,
   alsoletter={-},
   sensitive,%   
   basicstyle=\ttfamily\small,
   commentstyle=\slshape,%
   keywords={%
     tezos-node,%
     run,%
     generate, identity,%
     tezos-client,%
     gen, keys,bootstrapped,%
     rpc, get, post, list,%
     activate, account,with,%
     transfer, from, to,%
     wait, for, be, included,%
     originate, contract, transferring,%
     script,storage,code,%
     show, known,%
     running, %
   },%
   keywordstyle={\color{black}\bfseries},%
   morekeywords=[2]{%
     --rpc-addr,-l,%
     --init, --arg,--burn-cap, --fee%
   },%
   keywordstyle=[2]{\color{black}\slshape},%
   comment=[l]\#,%
   morestring=[d]",%"
   literate={tezos-client}{{\color{black}\bfseries t-c }}4%
   {tezos-client-full}{{\color{black}\bfseries tezos-client}}1%
}[keywords,comments,strings]%

\lstMakeShortInline[language=rpcs]|

\title{Introduction to the Tezos Blockchain}
\author{%
    \IEEEauthorblockN{
    Victor Allombert\IEEEauthorrefmark{1},
    Mathias Bourgoin\IEEEauthorrefmark{1} and
    Julien Tesson\IEEEauthorrefmark{1}
  }
  \IEEEauthorblockA{
    \IEEEauthorrefmark{1}
    Nomadic Labs, Paris, France \\
    Email: firstname.lastname@nomadic-labs.com \\
  }
}

\IEEEspecialpapernotice{HPCS 2019 INVITED TUTORIAL PAPER}

\newcommand\blfootnote[1]{%
  \begingroup
  \renewcommand\thefootnote{}\footnote{#1}%
  \addtocounter{footnote}{-1}%
  \endgroup
}

\usepackage{framed}

\begin{document}

\maketitle

\blfootnote{\begin{framed} {\small \noindent This is a draft paper of the tutorial submitted to HPCS 2019 (Dublin, Ireland from 15 to 19 July 2019). \\\url{http://hpcs2019.cisedu.info/4-program/tutorials-hpcs2019}} \end{framed}}

\begin{abstract}

  Tezos is an innovative blockchain that improves on several aspects
  compared to more established blockchains.
  It offers an original
  \emph{proof-of-stake} consensus algorithm and can be used as a
  decentralized smart contract platform. It has the capacity to amend
  its own economic protocol through a voting mechanism and focuses
  on formal methods to improve safety.

% Les blockchains c'est bien
% Les blockchains c'est le futur
% Les blockchains vont sauver le monde
% Les blockchains ne sont pas du HPC...
% ... mais elles vont sauver le monde
% du coup ils faut apprendre à s'en servir
\end{abstract}


\section{Introduction to Blockchains}

At the heart of blockchains lay  massively distributed and decentralized
programs that aim at bringing consensus (usually over a ledger) among
thousands of nodes.

Tezos~\cite{goodman2014tezos} is an innovative blockchain that
improves several aspects compared to more established blockchains like
Bitcoin~\cite{nakamoto2008bitcoin} or
Ethereum~\cite{wood2014ethereum}. It offers an original
\emph{'Proof-of-Stake'}~\cite{king2012ppcoin} consensus algorithm and
can be used as a decentralized smart contract platform. It has the
capacity to amend its own consensus algorithm (and more) through a
voting mechanism and focuses on formal methods to improve safety.

Tezos implementation is designed as a multi-layer software. A
\emph{peer-to-peer} layer ensures the connectivity with many other
nodes and passes the received messages to the next layer that sees this
network as a \emph{distributed database}. This layer pulls block
chunks from its neighbors, pushes new block identifiers, and passes
new information to the \emph{economic protocol} layer which implements
the consensus: it decides which blocks with which transactions
can be included in the ledger.

The economic protocol also embeds a smart contract platform. Smart
contracts are small programs that manage their associated tokens and
data storage and perform blockchain operations. Transactions in
the ledger can be more than just mere transfers of tokens (assets
residing on the blockchain), they can carry data and can be addressed
to a smart-contract. In this case they can be seen as a function call
which triggers the execution of the smart-contracts code.

To ensure the consistency of the ledger state, each block received by
a node has to be validated before the node agrees to transfer it to
its neighbor, thus it is very important for a smart contract execution
to be limited in both time and storage in order not to slow down the
network.

This tutorial aims at giving a broad presentation of what a blockchain
is and how to use and interact with Tezos.
% , and what are today's challenges for better scaling in terms of
% transactions per second.
We begin with a short introduction to blockchains in the rest of this
Section. Section 2 will detail some specifics of Tezos. The next
sections will focus on interacting with the Tezos blockchain through
multiple examples. Section 3 shows how to use the
blockchain through a client and \emph{via} a rich set of RPC calls
that can be used from any programming language. In Section 4, we
present how to use Tezos as a decentralized platform that can run
smart contracts. We present through examples how to build and run a
small contract while providing some technical details of the Tezos
smart contract programming language and platform.

\subsection{Blockchains building blocks}

Blockchains can be seen as an immutable database operating in a
decentralized network. They are built upon several key concepts and
tools:
\begin{itemize}
\item They use cryptography heavily to ensure users authentication as well as
  the database immutability
  % we only use hashes and digital signatures, not sure if it's heavy
  % we could actually say the contrary, "they use common and established cryptographic primitives, hash functions and digital signatures.
\item They offer a probabilistic solution to the \enquote{Byzantine
    generals problem}~\cite{lamport1982byzantine} for consensus among
  all participants (that we will call nodes in the rest of this
  tutorial) in the decentralized network
\item They use a peer to peer (P2P) gossip network for low-level
  communications between the nodes.
\end{itemize}
Thus blockchains are often called \emph{crypto-ledgers} as they can be
seen as an electronic book, recording transactions where users
identity and book immutability are cryptographically ensured.

In order to validate and append transactions to the ledger, all blockchains
follow a similar generic algorithm:
\begin{enumerate}
\item New transactions are broadcasted to all nodes which aggregate them
  in blocks.
\item The next block is broadcast by one or several nodes.
\item Nodes express their acceptance of a block by including its
  cryptographic hash in the next block they create.
\end{enumerate}

Blockchains face some common distributed systems challenges. To be
resilient to Sybil attacks\cite{douceur2002sybil} a solution is to
restrict the pool of block producers by tying it to the use of a
scarce resource. A difficulty is to choose this resource and create
incentives that push the majority of the network to be honest.
Restricting the pool of block producers can also lead to liveness
issues that have to be taken into account to make sure that the chain
does not stop whenever a block producer is offline. Blockchains also
have to consider malicious block producers and performance issues such
as network delays. Each blockchain provides its own set of solutions
to overcome these challenges but most of them rely on the foundations
set by Bitcoin. Blockchains also face the risk of forks: to
update their economic protocol, blockchains have to go through social
consensus and risk frictions in their user community that may lead to
the birth of \emph{hard forks} splitting the community in two parts
agreeing on two different chains. Such hard fork
occurred after the Ethereum DAO
hack\footnote{\url{https://www.bloomberg.com/features/2017-the-ether-thief/}}
leading to the birth of two blockchains: Ethereum Classic and Ethereum.

\subsection{Bitcoin: the electronic cash}

Bitcoin was introduced in 2008. % as the first known blockchain.
The main objective behind Bitcoin was to propose a decentralized
electronic cash system. Bitcoin is the name of the blockchain as well
as the name of the cryptocurrency it uses. Blockchains use tokens to
represent assets stored in the chain. Some of these tokens, as in
Bitcoin, can represent a currency. Bitcoin is also a decentralized
system (the actual blockchain) allowing users to store and transfer
their tokens. This system is the combination of a P2P network
associated with a consensus protocol to maintain consistency
between nodes. This consensus algorithm introduced the notion of
'Proof-of-Work' (PoW). With PoW, the scarce resource used to restrict
the pool of block producers (that Bitcoin calls 'miners') is computing
power. High computing power mostly demands very efficient computing
hardware associated with high energy consumption. The main idea is to
request that miners compete in the solving of a puzzle to earn the
right to produce (mine) the next block. The puzzle is built to be hard
to solve but easy to check. With PoW, Sybil attacks are difficult and
expensive. Associated with incentives to motivate miners to compete,
liveness is easy to achieve. In case of \emph{soft forks} (temporary
split) of the chain, which can be caused by bugs, network latency or
malicious mining, and to maintain consistency of the chain among all
the nodes, Bitcoin's consensus algorithm specifies to always keep the
longest chain between two forks. A chain is considered longer if its
total difficulty (summing the difficulty of the puzzles solved to mine
each block of the chain) is higher. The difficulty is adapted every
two weeks to maintain an average of $10$-minute intervals between two
consecutive blocks depending on the global computing power of the
miners. PoW is the solution currently used by most blockchains.

\subsection{Smart Contracts: decentralized platforms}

While Bitcoin focused on electronic cash, it also come with the concept
of decentralized computations: Bitcoin's script allows many
interesting forms of \emph{in-transaction} computation, and
% we could mention that bitcoin script already allowed many interesting forms of 'smart contracts', ethereum pushed the concept further.
others quickly
proposed to use blockchains to build decentralized computing
platforms.
This was popularized by the current
second-biggest blockchain, Ethereum, in 2014, which pushed the
concept further. The main idea is to see blockchains as
vending machines where users can pay for a service. In blockchain
platforms, these services are small programs living on the chain.
Users of the blockchain can store code in blocks and other users can
execute this code. These programs are called \emph{Smart Contracts}.
They can, of course, perform blockchain operations, such as token
transfers, but they can also be used for access control or to interact
with others. Some smart contracts, being fed with off-chain
information, serve as \emph{Oracles} selling trusted information.
Smart contracts can be used in many applications such as financial
contracts, developing new currencies on top of the basic
crypto-currency of the blockchain, voting systems, games or
crowdfunding.

%%% Local Variables:
%%% mode: latex
%%% TeX-master: "paper"
%%% TeX-engine: xetex
%%% End:

\section{Tezos specifics}

Tezos is an innovative blockchain which was presented in 2014.
Contrarily to most new blockchains at that time, it is not based on PoW.
In the following section, we present some distinctive features
of Tezos briefly such as its self-amending mechanism, its usage of a
'Proof-of-Stake' based consensus algorithm and its strong emphasis on
formal verification.

\subsection{Self amending blockchain}
Tezos is a self-amending crypto-ledger.
The protocol that validates blocks and implements the consensus
algorithm can amend itself.
Concretely a new protocol is downloaded from the network, compiled and
how-swapped to replace the current one.

The amendment procedure can be triggered in several ways, depending on
the protocol.
In the current Tezos economic protocol, an amendment can only be
triggered as the result of an on-chain voting procedure.
% It supports meta-upgrades: the
% protocol that guides how transactions are recorded on the chain can
% amend itself. % To achieve this deandes to abstract all operations of
% % a regular blockchain to keep amended protocols compatible with the
% % low-level layout responsible of the network communication (that we
% % call the shell). Thus, well known blockchains such as Bitcoin or
% % Ethereum can all be represented within Tezos by implementing the
% % correct interface to the shell. In Tezos, the generic operations of
% % regular blockchains are implemented as a purely functional module
% % abstracted into the shell. These operation are based on a generic data
% % structure representing any blockchain at an abstract level. The
% % blockchain is a concurrent data-structure represented by a shared
% % mutable singleton as a linked list of blocks of operations.
% We define self-amendment as the process used to upgrade the protocol
% over time through on-chain voting.

It helps to avoid forks of the chain
and reduces friction and splitting in the community. A protocol
amendment may consist of very important upgrades such as a switch to a
completely different consensus protocol. It can also consist in
smaller modifications such as extending the smart contract language,
modifying the rewarding system to enforce network participation or
adding new kinds of transactions (for instance adding anonymous ones). In
order to amend itself, Tezos uses an on-chain voting system where
users of the blockchain participate to propose, select, adopt or
reject new amendments. The voting process is currently divided in $4$
periods of \textasciitilde{}3 weeks (time is measured in blocks, Tezos
aiming at a $1$-minute intervals between two consecutive blocks):
\begin{enumerate}
\item Participants submit new protocol proposals (i.e. hashes of the
  protocol proposals source files)
\item A first vote selects a proposal among the submitted proposal
\item A side test chain spawns with the elected protocol
\item A final vote occurs that decides whether to upgrade(needing a supermajority
  of 80\% of positive votes)
\end{enumerate}

Tezos is built specifically in order to be self-amendable. Tezos nodes
are split into two parts: the \emph{protocol}, which is the self amendable
part, and which is isolated from the \emph{shell} that is responsible
for the low-level network
communications.

The shell can be written in any programming language. There can be
multiple implementations with different properties. It corresponds to
the first two layers of Tezos architecture.

The protocol has to run exactly the same way on all nodes. It is
responsible for validating blocks and operations. Operations, that are
aggregated into blocks, are what is stored in the ledger and are of
two kinds:
\begin{itemize}
\item ledger operations: transactions and origination (creation) of
  contracts
\item Proof-of-Stake operations such as endorsements or delegation that are
  described in Section~2.2.
\end{itemize}
The protocol can also trigger a protocol upgrade. In order to allow
all kinds of protocols to be compatible with the shell, Tezos reduces
the protocol interface as much as possible. In Tezos, the generic
operations of regular blockchains are implemented as a purely
functional module. Thus, well known blockchains such as Bitcoin or
Ethereum can all be represented within Tezos by implementing the
correct interface to the shell. The interface of the protocol is
primarily composed of two functions: \textbf{\texttt{apply}} and
\textbf{\texttt{score}}.

{$$\begin{array}{rcl}\text{\bf{\texttt{apply}}} & : & S \times B \to S \\
    \text{\bf{\texttt{score}}} & : & S \to \mathbb{N} \end{array}$$}

\par\noindent where $S$, the \emph{state} of the blockchain (the
ledger), is an immutable key-value store and $B$ is a block.
\textbf{\texttt{apply}} takes the current state and a block to produce
a new state. \textbf{\texttt{score}} computes the score of a state to
choose the preferred one between multiple states (implemented as the
longest chain in Bitcoin). A few other functions are exposed for
efficiency purposes, to document errors and provide protocol-dependent
RPCs.

Tezos protocols are written in the OCaml programming
language~\cite{leroy:hal-00930213}. OCaml is a powerful functional
programming language offering speed, an unambiguous syntax and
semantic, and a rich ecosystem that makes Tezos a good candidate for
formal proofs of correctness. To make it more resilient and less
error-prone, the protocol has only restricted access to the standard
library: for instance it uses no I/O functions, no threads, no unsafe
language traits. It also has access to specific libraries such as
formally verified cryptographic libraries or database abstractions.

%%% Local Variables:
%%% mode: latex
%%% TeX-master: "paper"
%%% TeX-engine: xetex
%%% End:

\subsection{'Proof-of-stake' based consensus algorithm}
The current Tezos protocol is based on a 'Liquid Proof-of-Stake'
(LPoS) consensus algorithm.

PoS is very different from PoW. It considers the stake (the number of
tokens) that users hold as the primary resource used to build the pool
of block producers (called \emph{bakers} in the Tezos ecosystem). In
the current Tezos consensus protocol, to push a block at a certain
level, bakers are randomly selected using a lazy infinite priority
list of baking slots. In order to participate in this random
selection, a baker must hold at least a roll of tokens (corresponding
to $10,000$ tokens\footnote{$10,000$ tokens while writing these lines.
  An amendment of the protocol, currently in the testing phase of the
  voting process described in the previous subsection, reduces the
  size of a roll to $8,000$ tokens}). The number of baking slots is
proportional to the number of rolls that a baker holds. However,
participants that do not hold enough tokens or who do not wish to bake
blocks can delegate their tokens to another baker, much like in Liquid
Democracy one can delegate its right to vote.
They keep the
ownership of their tokens but increase the stake of their delegate in
the random assignment of baking slots.
Delegation makes the PoS system more fair and participative and helps
balance a possible concentration of tokens in few hands.

In order to help the chain reach finality (the guarantee that a block
will not be revoked and that past transactions can never change)
faster, Tezos PoS mechanism introduces endorsements of baked blocks.
For each baked block, 32 endorsements (signatures) slots are created,
allowing chosen bakers to approve a block by signing it. Using endorsements,
the highest block resulting score is considered the head of the chain
where the score is:

$$\begin{array}{lcr}\text{\textbf{\texttt{score}}}(B_{n+1}) &=&%
1 + \text{\texttt{\textbf{score}}}(B_n) + \text{nb\_endorsements}\end{array}$$

In order to provide incentives to bakers for participating in the
network, the protocol rewards baking and endorsing. A baker earns
$16$ tokens for each block it bakes and $2$ tokens for each
endorsement it produces.

Two main malicious behaviors are also handled by the protocol:
\emph{double baking} and \emph{double endorsement}. A baker
perpetrates double baking when it injects two different blocks at the
same level. Double endorsements happens when a baker signs two
different blocks for the same level. The system punishes malicious
behaviors as follows: when a baker produces a block, a deposit bond of
$256$ tokens is frozen for $\sim\!\!\!\!2$ weeks ($64$ tokens for an
endorsement). During this period, if the baker/endorser is caught
cheating, the deposit and pending rewards (summing the rewards earned
baking and endorsing in the last $5$ cycles $-$ a cycle equals $4096$
blocks) of the cheater are forfeited.

Tezos LPoS consensus algorithm, \emph{via} its internal mechanism and
its associated incentives, solves the challenges presented in Section
1 without requiring significant computational power. It also focuses
on the users of the platform (instead of external actors as it is
possible with PoW): not only stake-holders can participate but all users
can, from large ones with many rolls to smaller ones that delegate
their stake.

%%% Local Variables:
%%% mode: latex
%%% TeX-master: "paper"
%%% TeX-engine: xetex
%%% End:

\subsection{Strong emphasis on formal verification}
Tezos uses as much as possible state-of-the-art programming languages
capacities to statically ensure the correctness of the implementation
and limit the possible runtime errors or attacks.

The code base is mainly written in the OCaml programming language,
whose robust static type system and memory management system rule out
many common runtime errors like null pointer exceptions or buffer
overflows.

%etrange phrase
Regarding cryptographic primitives implementation, whose importance in
terms of security is paramount for the blockchain, Tezos relies
on the HACL* library \cite{HACL:cryptoeprint} which is implemented in
Low* \cite{LowStar:ICFP} and extracted to C. The cryptographic
primitive implementation is formally proven to be \emph{memory safe},
\emph{functionally correct} and resistant to
\emph{side-channel attacks} at least at the level of C (secret independence of branching and
memory access).

Michelson, the Tezos \Sc{} language,  has been explicitly
designed to ease the readability and verifiability of contracts while
being low level enough to comply with the performance predictability
requirement of on-chain execution. The language is statically typed,
its formal semantics has been written in the Coq proof assistant
\cite{coq} and formal proofs of functional correctness of \Sc{}
using this semantics have been done.

%%% Local Variables:
%%% mode: latex
%%% TeX-master: "paper"
%%% TeX-engine: xetex
%%% End:

\section{Interacting with the Tezos Blockchain}
\label{sec:interact-with-tezos}

The architecture of Tezos is centered around two main components.

First, the \emph{node} (the corresponding executable file being called
{\tt tezos-node}) is responsible for connecting to peers through the
gossip network and updating the ledger's state (\emph{context}). As
all the blocks and operations are exchanged between nodes on the
gossip network, the node is able to filter and propagate data from/to
its connected peers. Using the blocks received on the gossip network,
the node keeps an up-to-date context. The node can be run with several
daemons such as {\tt tezos-baker-*} and {\tt tezos-endorser-*}
which take part of the consensus algorithm by, respectively, baking
and endorsing blocks.

Second, the \emph{client} ({\tt tezos-client}) is the main tool to
interact with a Tezos blockchain node.

There are currently $3$ public Tezos networks:
\begin{itemize}
\item |mainnet| which is the current incarnation of the Tezos
  blockchain. It runs with real tez (Tezos tokens) that have been baked or
  allocated to the donors of July 2017 fundraiser. It has been live and open
  since June 30th 2018
  \item |alphanet| which is based on the §mainnet§ code base but uses
free tokens. It is the reference network for developers wanting to
test their software.
  \item |zeronet| which is the testing network, with free tokens and
frequent resets.
\end{itemize}

In this tutorial, we will use the §alphanet§ Tezos network.

In the following sections, we assume that the reader have access to the Tezos binaries. A pre-configured Tezos environment can be found in the provided virtual machine. Otherwise, it is possible to install a Tezos environment from source (using the ocaml package manager (opam) and compiling from source) or with docker (using scripts and images).
All the instructions to install and run the Tezos software from source or from docker can be found at \url{http://tezos.gitlab.io/master/introduction/howtoget.html}.

\subsection{Seting up a Tezos node (\textbf{demo})}
\label{sec:set-up-node}

The \emph{node} ({\tt tezos-node}) can be considered as the
\emph{access point} to the Tezos blockchain and stores all the data
necessary to run the blockchain. In practice, the node's data is
stored (by default) into the § ~/.tezos-node§ directory.

To be connected to the network, a node must have a proper network
identity to be globally identified.

To generate an identity, the following command should be run:
  \begin{lstlisting}[language=rpcs]
    tezos-node identity generate
  \end{lstlisting}

The generated identity will be stored as a pair of cryptographic keys
that are used by the node to send encrypted messages, but it is also
used as an antispam measure (to prevent Sybil attacks) based on a
lightweight PoW.

When the identity is generated, we can run a node using:
  \begin{lstlisting}[language=rpcs]
  tezos-node run --rpc-addr 127.0.0.1
\end{lstlisting}

The {\tt --rpc-addr 127.0.0.1} argument is used to allow
communications with clients on the local host only. The node is now
able to connect to the Tezos network and will start its
\emph{bootstrap} phase. It consists in downloading all the blocks from
the chain using the distributed network. This procedure can be very
long as the chain data is growing invariably every day. To speedup the
process (from days to minutes), it is possible to start a node from
a snapshot of the chain\footnote{To avoid to download a fake chain, it
  is necessary to carefully check that the block hash of the imported
  block is included in the chain. However, we do not detail the
  procedure here.} by running:
  \begin{lstlisting}[language=rpcs]
    tezos-node snapshot import last.full
\end{lstlisting}

This command is able to read all the necessary data stored in the {\tt
last.full} file, validate it and import it in the node storage. The
imported data consist in a partial ledger state (that can be
reconstructed on request) and all the blocks of the chain since the
genesis. It is also possible to set up a lightweight node targeting
low resource architectures by running a partial chain using a
\emph{rolling} snapshot. When the import is done, one can run the
node, and wait a few minutes to download the new blocks spawned since
the snapshot file was exported.

\subsection{Using basic client commands (\textbf{demo})}
\label{sec:using-basic-client}

The \emph{client} (§tezos-client-full§) is a user-friendly interface
that can be used to interact with a node. As it is based on
\textsc{JSON RPCs}, it can be requested by various third-party
applications. For the sake of brevity, we will use §tezos-client§
instead of §tezos-client-full§ in the rest of the document.

The client can be used to check if the current head of the local node
is up-to-date using §tezos-client bootstrapped§. This command will
hang and return only when the node is synchronised.

The client is also able to handle a simple wallet, stored (by default)
in the § ~/.tezos-client§ directory. It mainly contains $3$ files
: {\tt public\_keys}, {\tt secret\_keys} and {\tt public\_key\_hashes}
(Tezos addresses : \emph{tz1}). To generate a new pair of keys to be
used locally for \textbf{Bob}, we can run:
\begin{lstlisting}[language=rpcs]
  tezos-client gen keys bob
\end{lstlisting}

In order to test the network and help users get familiar with the
system, one can obtain free tokens from a faucet\footnote{Please drink
  carefully and don't abuse the faucet: it only contains 30,000
  wallets for a total amount of 760,000,000 tokens.}:
\url{https://faucet.tzalpha.net/}. This service will provide a simple
wallet formatted as a \textsc{JSON} file. The account can be activated
for an identity using:
\begin{lstlisting}[language=rpcs]
  tezos-client activate account alice with "tz1__xxxxxxxxx__.json"
\end{lstlisting}

We can now check the balance of this account using:
\begin{lstlisting}[language=rpcs]
  tezos-client get balance for alice
\end{lstlisting}

It is time to try to transfer some tokens from one account to another.
To transfer $1$ token from {\tt Alice}'s account to {\tt Bob}'s one,
we can run
\begin{lstlisting}[language=rpcs]
  tezos-client transfer 1 from alice to bob --fee 0.05
\end{lstlisting}

The {\tt --fee} argument stands for the fees associated to an
operation in order to encourage bakers to include our operation in a
block. To be sure that the operation is well included in the chain, it
is advised to wait $60$ blocks (\textasciitilde{}$60~$min) to
consider it as a \emph{valid transaction}:
\begin{lstlisting}[language=rpcs]
  tezos-client wait for <operation hash> to be included
\end{lstlisting}

Client commands are high-level operations implemented using the set of
RPCs exposed by the Tezos node. The next section presents how the
transfer operation can be implemented manually using some of these
RPCs.

\subsection{Using RPCs}

In this section, we show how to transfer tokens from one account to
another by using RPCs. We will use the client to call the RPCs of the
associated node.

The whole set of RPCs can be found in the \emph{JSON/RPC interface}
section of the online Tezos documentation \cite{tezosdoc} or by using
the following client command:
\begin{lstlisting}[language=rpcs]
  tezos-client rpc list
\end{lstlisting}

The \texttt{-l} option of the client logs all the requests to the node.
The following command shows all the RPC calls made during a transfer.
\begin{lstlisting}[language=rpcs]
  tezos-client -l transfer 1 from bob to alice --fee 0.05
\end{lstlisting}

When executed, we can see that a simple transfer consists of around
$20$ calls to the node.

In this tutorial, we will only focus on the $10$ mandatory calls to
make a transfer. For readability we will use some shortcuts.

\begin{itemize}
\item |BOB| corresponds to Bob's public key.
\item |ALICE| corresponds to Alice's public key.
\item |HEAD_HASH| corresponds to the hash of the block head.
\item |CHAIN_ID| corresponds to the id (hash) of the chain.
\end{itemize}

\begin{enumerate}

\item In Tezos, account operations are numbered, in order to prevent
  replay attacks. Nodes can be queried to get the current counter and
  compute a new counter (by incrementing the current one) to forge a
  new operation. If the new operation has an incorrect counter, it can
  be ignored, or delayed. The following command gives the current
  counter for Bob's account.
\begin{lstlisting}[language=rpcs]
  tezos-client rpc get
    /chains/main/blocks/head/context/contracts/BOB/counter
\end{lstlisting}

\item For signature check of the incoming transaction, it is mandatory
  to verify that the sender public key is known on the blockchain.
  \begin{lstlisting}[language=rpcs]
tezos-client rpc get
    /chains/main/blocks/head/context/contracts/BOB/manager_key
  \end{lstlisting}

\item To make sure that the transaction will be
  added into the blockchain, we have to make sure that the node is
  bootstrapped (ie. that it is synchronized with the other nodes in
  the system).
  \begin{lstlisting}[language=rpcs]
    tezos-client rpc get /monitor/bootstrapped
  \end{lstlisting}

\item Some values have to be given to the transaction operation, in
  particular the |gas_limit| and
  |storage_limit| (see Sec~\ref{sec:limit-exec-time} and \ref{sec:limit-data}) %todo il faut expliquer ce que c'est.
  constants can be queried :
  \begin{lstlisting}[language=rpcs]
    tezos-client rpc get
    /chains/main/blocks/head/context/constants
  \end{lstlisting}
  The needed information can be extracted from the JSON answer:
  \begin{lstlisting}[language=rpcs]
    { ...
      "hard_gas_limit_per_operation": "400000",
      ...
      "hard_storage_limit_per_operation": "60000" }
  \end{lstlisting}

\item The hash of the head is also needed:
  \begin{lstlisting}[language=rpcs]
    tezos-client rpc get /chains/main/blocks/head/hash
  \end{lstlisting}

\item As well as the id of the chain:
  \begin{lstlisting}[language=rpcs]
    tezos-client rpc get /chains/main/chain_id
  \end{lstlisting}

\item We can now simulate the execution of our operation:
  \begin{lstlisting}[language=rpcs]
    tezos-client rpc post /chains/main/blocks/head/helpers/scripts/run_operation
  \end{lstlisting}
  Here we
  use a |POST| that demands a JSON input.
  \begin{lstlisting}[language=rpcs]
    { "branch": "HEAD_HASH",
      "contents":
      [ { "kind": "transaction",
        "source": "BOB",
        "fee": "50000",
        "counter": "4",
        "gas_limit": "400000",
        "storage_limit": "60000",
        "amount": "1000000",
        "destination": "ALICE" } ],
      "signature": ANY_SIGNATURE ... }
  \end{lstlisting}
    We can use |ANY_SIGNATURE| to make the simulation without signature checks.
    In the JSON answer, we get how much gas and storage were consumed.
    |{... "consumed_gas": "10100" ...}|

  \item We can now adjust the fees, gas limit and storage limit based on
    the last RPC result and run the simulation with signature check.
    \begin{lstlisting}[language=rpcs]
      tezos-client rpc post
      /chains/main/blocks/head/helpers/preapply/operations
    \end{lstlisting}

    \begin{lstlisting}[language=rpcs]
      [ { "protocol": "ProtoALphaALphaALphaALphaALp...",
        "branch": "HEAD_HASH",
        "contents":
        [ { "kind": "transaction",
          "source": "BOB",
          "fee": "1269",
          "counter": "1",
          "gas_limit": "10200",
          "storage_limit": "0",
          "amount": "1",
          "destination": "ALICE" } ],
        "signature": "edsigtf12Ls...} ]
\end{lstlisting}

  \item We can now inject the operation:
    \begin{lstlisting}[language=rpcs]
      tezos-client rpc post
      injection/operation?chain=main
    \end{lstlisting}
    This RPC call take an hex-encoded
    signed operation as input (|"09115800..."|) and returns an
    operation identifier (|"opDerPd..."|).

  \item An additional |POST| RPC call (that is not used by the
    client) can be helpful to compute the hex-encoded operation:
    \begin{lstlisting}[language=rpcs]
      tezos-client rpc post
      /chains/main/blocks/head/helpers/forge/operations
    \end{lstlisting}

\end{enumerate}

%%% Local Variables:
%%% mode: latex
%%% TeX-master: "paper"
%%% TeX-engine: xetex
%%% End:

\section{Tezos as a decentralized platform}
\label{sec:tezos-as-decentr}
\newcommand{\tzlines}[4][last] {\lstinputlisting[breaklines=true,language=michelson,numbers=left,numberstyle=\tiny,numbersep=3pt,firstnumber=#1,firstline=#3,lastline=#4]{#2}}
\newcommand{\votelines}[3][last] {\tzlines[#1]{vote_simpl}{#2}{#3}}
\newcommand{\voteline}[2][last] {\votelines[#1]{#2}{#2}}
\newcommand{\oraclelines}[3][last] {\tzlines[#1]{oracle}{#2}{#3}}
\newcommand{\oracleline}[2][last] {\oraclelines[#1]{#2}{#2}}
\newcommand{\insurancelines}[3][last] {\tzlines[#1]{insurance}{#2}{#3}}
\newcommand{\insuranceline}[2][last] {\insurancelines[#1]{#2}{#2}}

As mentioned before, Tezos economic protocol not only handles a
registry of transactions, but also has support for \Sc{}s.

\SC{}s are small programs registered in the blockchain together with a
private data storage: meaning that only the contract can interact with
the storage, but the data are publicly visible. A contract registered
in the chain is said to be \emph{originated} and it has an address
prefixed by KT1 which is given in the contract's origination block.

They are executed by performing specific transactions to their
associated account.
The transaction carries data that are passed as a program
parameters and can thus be viewed as a procedure call.

The execution of a \Sc{} can change the state of its storage and
trigger on chain transactions.

\SC{} languages are usually Turing-complete. However the
replicated nature of the contract storage and the liveliness
requirement of the consensus algorithm imposes some restrictions on
their execution.

\subsection{Limited execution time}
\label{sec:limit-exec-time}
Any call to a smart contract, once included in a block, will be
executed on every node in the P2P network, because they have to
validate the block before including it in their view of the chain and
before passing it to their neighbors. It means that the execution
time of each smart contract call included in a block has to fit
multiple times in the inter-block time of the chain (1 minute for
Tezos)  to ensure its liveliness.

Thus each call is allowed a bounded quantity of computation:
the \Sc{}s interpreter uses the concept of \emph{gas}.
Each low-level instruction evaluation burns an amount of gas which is
crafted to be proportional to the actual execution time and if an
execution exceeds its allowed gas consumption, it is stopped
immediately and the effects of the execution are rolled back. The
transaction is still included in the block and the fees are taken, to
prevent the nodes from being spammed with failing transactions.

In Tezos, the economic protocol sets the gas limit per block and for
each transaction, and the emitter of the transaction also set an upper
bound to the gas consumption for its transaction. The economic
protocol does not require
the transaction fee to be proportional to the gas upper bound,  however
the default strategy of the baking software (that forges blocks)
provided with Tezos current implementation does require it.

\subsection{Data storage}
\label{sec:limit-data}
Each \Sc{} on the chain possesses its own storage, only accessible to
the contract. As this storage is replicated on every node that runs the
chain, it has to be of limited size in order to avoid that the chain
context grows out of control. A cost is set for storage allocation
(currently 0.001\tz{} per byte) to restrain storage usage.

\subsection{Michelson: Tezos' smart-contract programming language}
\label{sec:mich-tezos-smart}

\subsubsection{Design rationale}
\label{sec:design-rationale}
The constrained context in which \Sc{}s operate imposes strong contradicting
constraints on the language design.

Because we need to be able to accurately account for resources
consumption, the language has to be interpreted. The interpreter is
thus counting gas at each ``opcode'', and each opcode cost has to be
fairly simple to guess.  This tends to push to a low-level language,
at the same time, however, the resource constraint will lead people to
write their program in this language. Indeed, they do not want to rely
on a very high-level language with a compiler performing many
under-the-hood transformations, preventing cost predictions.
Therefore, the
language has to be high-level enough to be programmable by a human.

Furthermore, as the program will be stored on chain in this language,
it is of paramount importance that they can be audited easily. The
language has to be simple, high-level enough and should offers as few
 means of code obfuscation as possible in order not to mislead the
reader.

One more constraint is that the language gives as much guarantees as
possible statically, as once published on the chain, it is not
possible to modify it to correct bugs anymore. So we want to
have a strong type system that prevents as much runtime error as
possible.

\subsubsection{A stack language with high-level data structures}
Michelson, the \Sc{} language on Tezos is a stack based
language \emph{à la} Forth with strict static type checking and
high-level data structures \emph{à la} ML.

A Michelson program is a sequence of instructions which
modify the stack given as a program input. The initial stack contains
only the calling argument and the contract's storage, and the program
must end with a stack containing only a list of operations paired with
the new value of the storage.

This led to a rather simple interpreter, with simple cost model for
most operations, but with high-level data structures (such as maps,
sets, lists and algebraic data types) to help the writing of \Sc{}s.

The operations -- i.e. Tezos transactions (including calls to other
contracts), contract creations and delegate setting -- will only be
executed after the program returns. This prevents reentrancy bugs
(which are hard to spot and have costs millions and provoked the hard
fork on Ethereum after the DAO attack).
We will discuss contract interaction with an example hereinafter
(\ref{insurance}).

The type of each instruction describes the states of the stack before
and after the instruction.
For example, the instruction !DUP! has type !'a:'S -> 'a:'a:'S! meaning
that when starting from a stack whose top element has type !int!,
the duplication of the top element leads to a stack with one more
element of type !int! on top of it (i.e. !int:int:'S!).
Thus the type checking of the contracts ensure that no instruction can
failed because of a malformed stack.

While the type of the Michelson instruction is polymorphic, the type
of contracts arguments and storage have to be monomorphic. This is
partly to keep the type checking simple enough to be done efficiently:
contract type checking consumes gas and has to be efficient.

Rather than going into the details of all the language instructions,
we will provide here two programs examples. The interested reader can
find the full description of the language in the Tezos documentation
\cite{michelson-doc}.

\subsubsection{A voting contract}
\label{sec:voting-contract}

As a first example, we will describe a voting contract. The use cases
for such a contract range from voting for your favourite supercomputer in
a TV show, to  registering vote for important decisions in a
decentralized infrastructure.
The first one is a bit simpler to implement than the second as we
don't have to check the identities of the voters, so we will focus on
this: an open vote with a fee, to determine the preference of
the voters in a fixed list of choices.

In the following lines, we will present an abstraction of the state of
the stack with a comment (prefixed by !#!) after each relevant block of code.

We start with a storage which holds the names that voters are
allowed to vote for, associated with the number of votes they received.
We start our program by declaring the type of the storage:
a map from !string! to !int!.

\voteline{1}
Then we specify the type of the parameter:
\voteline{3}
and now we can write the code of the contract.
The contract execution starts with the parameter paired on top of an
empty stack:
\votelines{4}{5}
First we verify that the caller send us enough token to be able to
vote. If not, we make the call fail.
\votelines{6}{12}
 !AMOUNT! pushes on the stack the
number of tokens received from the contract caller, !PUSH 'a cst! pushes
the given constant !cst! of type !'a! on the stack.
!IFCMPGT! is a macro which  compares the two
numbers on top of the stack (removing them in the process) and  if the
first element was greater it executes its first parameter, otherwise the
second.

If payment is sufficient, we prepare the stack by duplicating the pair
holding the parameter and the storage, the first (name, map) pair will
be consumed by !GET! to obtain the current number of votes, while the
second will be used to produce the new map.

\votelines[12]{13}{14}
To get the value of interest from the storage we first destruct the
pair to get a stack with the key on top and the map beneath, and then
we apply the !GET! instruction.
\votelines[14]{16}{17}
We get the current count for the voted name or !None! if the key was
not in the map. If the count is some integer value, we add $1$ to this
value, if not we fail because the vote is for an unknown name.
\votelines{19}{22}
We now reorder the elements to prepare the stack to use !UPDATE!
in order to  update the map with the new count. !DIP! allows to work
on the element below the stack top, !SWAP! exchanges the two top elements
of the stack.
\votelines{24}{27}
We get a new storage, that we pair with an empty list of operations to
match the return type of the contract, a pair (list of operations, storage).
\votelines{28}{30}

This rather simple program can be extended in many ways. For example, a
deadline for the vote could be fixed by storing the end date in the
storage and comparing it to the value pushed by !NOW! on the stack.
Or we could grant the right to add new names to the map for voters
transferring a bigger amount to the  contract.
Finally, we could store the addresses of voters (obtained
with the instruction !SENDER!) and reward the voters who voted for the
winner.

All these improvements are left as exercises for the interested reader.

\subsubsection{Inter-contract calls}
\label{insurance}
As stated before, to prevent reentrancy bugs, calls to other contracts
are performed at the end of the current contract execution, thus to
emulate a procedure call expecting a return data, inter-contract calls
will have to use callbacks.

Let us take as an example an insurance contract $A$, which calls a
meteorological oracle contract $B$, with a given parameter like a date
to obtain a related data (say the hydrometry of the given date).
The contract $A$ will pass to $B$ a callback identifier, like the
insured address, $B$ will now call $A$ with the relevant data  and the
callback identifier.
The contract $A$ can now proceed to the refunding of the legitimate
contracting party depending on the data received from the oracle.

We start with a simple oracle  which is a contract that just
encapsulates a map but guarantees that only a trusted source -- whose
address is in the contract storage -- can modify the map.

For the sake of conciseness, we omit in this contract the usual code
allowing the oracle manager, whose key is stored in the contract,  to
withdraw the tokens held by the contract.
As contract will probably be non-spendable in the future, meaning the
owner of the contract will not be able to transfer tokens of the
contract, only code will, this would
freeze the tokens associated to the contract.

The parameter given to the contract is either a new key-data pair to
be updated, or a request for the data matching a given key.
\oraclelines{1}{4}
and the storage is a map together with the manager key:
\oraclelines{5}{9}

The code separates the parameter from the storage and then
checks with !IF_RIGHT! whether the call is an update (right of the
!or! type) or a client query (left of the !or!).
\oraclelines{11}{15}
For data update, we first check that the data are indeed provided by
the registered manager, and fail if it is not the case:
\oraclelines{17}{23}
The map is then updated with the provided values:
\oraclelines{24}{27}

If the call is a request from a client we retrieve the data from the
map and then craft a call to the sender with this value as parameter.
\oraclelines{29}{48}
% - Oracle

We can now define our insurance contract. Its parameter type has to
be the type expected by the oracle (The one used in the oracle
transfer).
In this small example it is
easy, but for more general-purpose oracle interacting with more general
purpose client contract, the later won't be able to have all the same type,
so the usual mechanism is to originate a proxy contract whose type
satisfies the oracle requirement and which can relay the calls between
the oracle and the client contract.

We will assume here that the insurance contract is issued for a
one-time insurance:  given a rain level threshold at a
certain point in time, it will redeem one or the other of the
registered addresses of the contracting parties.

The contract can be called by anyone with !Right Unit! to trigger the
redeeming mechanism or by the oracle with !Left (Some level)! (callback).
\insurancelines{1}{3}

The storage holds the contract parameters:
timestamp at which the rain level should be checked,
rain level threshold, redeeming addresses and address of the oracle to
consult.
\insurancelines{4}{17}

The code inspects the given parameter, if the parameter
is !Left (Some level)! then we first check that the sender is indeed
the oracle and then proceed to the redeeming:
\insurancelines{18}{32}
Else the call triggers the call to the oracle.
\insurancelines{34}{41}

\subsection{Contract origination and call}
\lstset{language=rpcs}
To originate the voting contract for a vote on your favorite
supercomputer, we can use Alice's account with the following command:
\begin{lstlisting}
tezos-client originate contract vote\
  for alice transferring 0 from alice \
  running ./vote.tz \
  --init\
  '{ Elt "Sierra" 0 ; Elt "Summit" 0 ;
    Elt "Sunway" 0 ; Elt "Tianhe-2A" 0 }'\
  --burn-cap 1
\end{lstlisting}
The elements of the storage have to be in alphabetical order.

Then we can vote for Summit, using the following transaction:
\begin{lstlisting}
tezos-client transfer 0.005 from bob to vote\
  --arg '"Summit"' --burn-cap 1
\end{lstlisting}

If we issue the transaction with a 0.001 instead of 0.005, the
transaction will fail, so we will keep the 0.001\tz{} but we will
loose the fees for the baker.

To test our Oracle/insurance example, we first originate the oracle,
as we need to initialise the insurance storage with the  KT1 address
of the oracle. This address is generated when the contract is
originated.

The initial storage use the address of Alice, tz1\_XXXX, as oracle
manager, meaning that only transactions initiated by Alice can add
data to the contracts map.

\begin{lstlisting}
tezos-client originate contract oracle\
  for alice transferring 0 from alice\
  running ./oracle_ok.tz\
  --init 'Pair { } "tz1_XXXX" ' --burn-cap 1
\end{lstlisting}

The origination receipt gives us the contract address, or we can
retrieve it later with the client:
\begin{lstlisting}
tezos-client show known contract oracle
\end{lstlisting}
Let say the address of the contract is KT1\_YYYY, we now can
originate our insurance:

\begin{lstlisting}
tezos-client originate contract insurance \
  for bob transferring 100 from bob\
  running ./insurance.tz --init\
  'Pair
    (Pair "2019-05-07 23:22:25+00:00"
      (Pair (Pair "tz1_AAAA" "tz1_BBBB") 10))
    KT1_YYYY'
\end{lstlisting}

Alice's account can feed the oracle contract with data:
\begin{lstlisting}
tezos-client transfer 0 from bootstrap1 to oracle
  --arg 'Right
     (Pair "2019-05-07 23:22:25+00:00" 15)'
\end{lstlisting}

and we can check that the data is indeed in the storage  of the
contract by inspecting it:
\begin{lstlisting}
tezos-client get script storage for oracle
\end{lstlisting}

Finally anyone can trigger our insurance:
\begin{lstlisting}
tezos-client transfer 0 from charlie to insurance
  --arg 'Right Unit' --burn-cap 1
\end{lstlisting}

The receipt shows that:
\begin{itemize}
\item the insurance contract makes a transfer to the oracle
\item the oracle makes a transfer back to the insurance contract
\item the insurance contract makes a transfer to the registered
  contracting address
\end{itemize}

%%% Local Variables:
%%% mode: latex
%%% TeX-master: "paper"
%%% TeX-engine: xetex
%%% End:

\section*{Glossary}
\balance
\let\descrold\descriptionlabel
\renewcommand{\descriptionlabel}[1]{\descrold{#1:}}
\begin{description}
\item [Baker] entity responsible for selecting operations to produce a block in Tezos
\item [Block] set of operations, aggregated in the blockchain
\item [Blockchain] distributed database formed as a list of blocks
\item [Client] entity responsible for interacting with a node
\item [Context] Ledger's state (accounts balance, contracts, \ldots)
\item [Cycle] set of consecutive blocks
\item [Delegate] entity to which an account has delegated stake
\item [Endorser] seal of approval for a block
\item [Liveness] mandatory property allowing the system to progress
\item [Miner] entity responsible for selecting operations to produce a block
\item [Node] entity responsible for connecting to a Tezos network
\item [Operation] transforms the context
\item [Oracle] off-chain third party that can deliver data
\item [Origination] operation to create an account that can contains a contract or be delegated
\item [PoS] Proof-of-Stake
\item [Pow] Proof-of-Work
\item [Roll] amount of tokens used to determine delegates' rights
\item [RPC] Remote Procedure Call
\item [Self-amending] ability to update itself seamlessly
\item [Smart contract] originated account which is associated to a Michelson script
\item [Stake] amount of token
\item [Storage] blockchain data necessary to run a node
\item [Sybil attack] take over the network by flooding malicious identities
\item [Token] unit of value 
\item [Tz1] Tezos implicit account address
\item [KT1] Tezos originated account address
\end{description}

%%% Local Variables
%%% mode latex
%%% TeX-master "paper"
%%% TeX-engine xetex
%%% End

\bibliography{biblio}

\begin{thebibliography}{10}
\providecommand{\url}[1]{#1}
\csname url@samestyle\endcsname
\providecommand{\newblock}{\relax}
\providecommand{\bibinfo}[2]{#2}
\providecommand{\BIBentrySTDinterwordspacing}{\spaceskip=0pt\relax}
\providecommand{\BIBentryALTinterwordstretchfactor}{4}
\providecommand{\BIBentryALTinterwordspacing}{\spaceskip=\fontdimen2\font plus
\BIBentryALTinterwordstretchfactor\fontdimen3\font minus
  \fontdimen4\font\relax}
\providecommand{\BIBforeignlanguage}[2]{{%
\expandafter\ifx\csname l@#1\endcsname\relax
\typeout{** WARNING: IEEEtran.bst: No hyphenation pattern has been}%
\typeout{** loaded for the language `#1'. Using the pattern for}%
\typeout{** the default language instead.}%
\else
\language=\csname l@#1\endcsname
\fi
#2}}
\providecommand{\BIBdecl}{\relax}
\BIBdecl

\bibitem{goodman2014tezos}
L.~Goodman, ``Tezos -- a self-amending crypto-ledger,''
  \url{https://www.tezos.com/static/papers/white\_paper.pdf}, 2014.

\bibitem{nakamoto2008bitcoin}
S.~Nakamoto, ``{Bitcoin: A Peer-to-Peer Electronic Cash System},''
  \url{https://bitcoin.com/bitcoin.pdf}, 2008.

\bibitem{wood2014ethereum}
G.~Wood, ``Ethereum: A secure decentralised generalised transaction ledger,''
  \url{http://gavwood.com/paper.pdf}, 2014.

\bibitem{king2012ppcoin}
S.~King and S.~Nadal, ``Ppcoin: Peer-to-peer crypto-currency with
  proof-of-stake,'' \emph{self-published paper, August}, vol.~19, 2012.

\bibitem{lamport1982byzantine}
L.~Lamport, R.~Shostak, and M.~Pease, ``{The Byzantine generals problem},''
  \emph{ACM Transactions on Programming Languages and Systems (TOPLAS)},
  vol.~4, no.~3, pp. 382--401, 1982.

\bibitem{douceur2002sybil}
J.~R. Douceur, ``{The Sybil Attack},'' in \emph{International workshop on
  peer-to-peer systems}.\hskip 1em plus 0.5em minus 0.4em\relax Springer, 2002,
  pp. 251--260.

\bibitem{leroy:hal-00930213}
\BIBentryALTinterwordspacing
X.~Leroy, D.~Doligez, A.~Frisch, J.~Garrigue, D.~R{\'e}my, and J.~Vouillon,
  ``{The OCaml system release 4.07: Documentation and user's manual},''
  {Inria}, Intern report, Jul. 2018. [Online]. Available:
  \url{https://hal.inria.fr/hal-00930213}
\BIBentrySTDinterwordspacing

\bibitem{HACL:cryptoeprint}
J.~K. Zinzindohou{\'{e}}, K.~Bhargavan, J.~Protzenko, and B.~Beurdouche,
  ``Hacl*: A verified modern cryptographic library,'' Cryptology ePrint
  Archive, Report 2017/536, 2017, \url{https://eprint.iacr.org/2017/536}.

\bibitem{LowStar:ICFP}
\BIBentryALTinterwordspacing
K.~Bhargavan, A.~Delignat{-}Lavaud, C.~Fournet, C.~Hritcu, J.~Protzenko,
  T.~Ramananandro, A.~Rastogi, N.~Swamy, P.~Wang, S.~Z. B{\'{e}}guelin, and
  J.~K. Zinzindohou{\'{e}}, ``Verified low-level programming embedded in
  {F*},'' \emph{CoRR}, vol. abs/1703.00053, 2017. [Online]. Available:
  \url{http://arxiv.org/abs/1703.00053}
\BIBentrySTDinterwordspacing

\bibitem{coq}
{The Coq Development Team}, ``{The Coq Proof Assistant},''
  \url{http://coq.inria.fr}.

\bibitem{tezosdoc}
``{The Tezos Developer Resources},'' \url{http://tezos.gitlab.io/master/}.

\bibitem{michelson-doc}
{The Tezos Development Team}, ``{Michelson Reference Manual},''
  \url{http://tezos.gitlab.io/master/whitedoc/michelson.html}.

\end{thebibliography}
\bibliographystyle{IEEEtran}
% \section{Biographies}

\begin{IEEEbiography}{Victor Allombert} obtained his PhD in Computer
  Science from the University of Paris-Est Créteil in 2017. He is
  currently a research engineer at Nomadic Labs.

  His research areas are High Performance Computing applied to
  structured parallel programming languages and formal approaches for
  designing programming languages. During his PhD thesis, he designed
  and implemented a functional language for programming hierarchical
  architectures: MultiML.

  Web site: \url{https://www.lacl.fr/vallombert/}
\end{IEEEbiography}

\begin{IEEEbiography} {Mathias Bourgoin} obtained his PhD in Computer
  Science from the University Pierre et Marie Curie of Paris in
  2013. He is currently tenured Assistant Professor at the University
  of Orléans on extended leave and Senior Research engineer at Nomadic
  Labs.  His research interest are efficient abstractions for
  heterogeneous programming. In particular, he explores the use of
  high-level statically typed languages to design libraries combined
  with domain specific languages that make high performance
  heterogeneous programming (mostly using GPUs) simpler and safer.
\end{IEEEbiography}

\begin{IEEEbiography} {Julien Tesson} obtained his PhD in Computer
  Science from the University of Orléans in 2011. He is currently
  tenured Assistant Professor at Université Paris-Est Créteil on
  extended leave and Senior Research Engineer at Nomadic Labs. His
  research interests include the semantics and implementation of
  parallel languages and the verification of programs.
\end{IEEEbiography}

\begin{IEEEbiography}{Nomadic Labs} expertise centers around the research and
  development of products and services in various domains of computer
  science, namely distributed, decentralized, and formally verified
  systems. It is the main contributor of the Tezos blockchain.
\end{IEEEbiography}
\end{document}